\documentclass[%
reprint, 
superscriptaddress, 
showpacs,
amsmath, amssymb, 
aps, 
prb,
]{revtex4-1}
\usepackage{graphicx}
\usepackage{float}
\usepackage{dcolumn}
\usepackage{bm}
\usepackage{hyperref}
\hypersetup{colorlinks=true, citecolor=blue, urlcolor=blue, linkcolor=blue}
\begin{document}
	\title{Non-Hermitian total-loss high-order topological insulator based on 1D Su-Schrieffer-Heeger (SSH)}
	\author{Hui-Chang Li} \author{Jing-Wei Xu} \author{Chen Luo} \author{Tai-Lin Zhang} \author{Jian-Wei Xu} \author{Xiang Zhou} \author{Yun Shen} 
	\email{shenyun@ncu.edu.cn}
	\affiliation{Department of Physics, Nanchang University, Nanchang 330031, China}  
	\author{Xiao-Hua Deng} \email{dengxiaohua0@gmail.com}
	\affiliation{Institute of Space Science and Technology, Nanchang University, Nanchang 330031, China}
	

	\date{\today}
	\begin{abstract}
		
		Non-Hermiticity alters topology with the presence of non-Hermitian factors in topological systems. Most existing non-Hermitian topological systems derive their topological phases from Hermitian components, that is, the gain and loss of the system are considered simultaneously. In this work, we reveal two-dimensional non-Hermitian high-order topological insulator based on one-dimensional SSH chain, the nontrivial topology of which induced by total-loss. By introducing the loss of a specific configuration, we get a band gap with corner and edge states whose topology is characterized by the gapped wannier band. In addition, we demonstrate the existence of 'real-energy' edge states in pseudo-PT symmetric domain wall system. These results can be easily implemented in experiments, and promote the research of topological transmission of lossy systems in the real world.
		
	\end{abstract}
	\maketitle
	
	\section{\label{1}Introduction}
	
	Hermiticity of system is the basic assumption of quantum mechanics, which plays an important role in the study of topological states and allows the existence of real eigenvalues and orthogonal eigenstates which help to define the classification of topological phase~\cite{1-1,1-2,1-3} and various topological invariants~\cite{2-1,2-2,2-3,2-4,2-5,2-6,2-7}. Nowadays, the research of topological states~\cite{3-1,3-3} has been extended to various fields, such as superconductor~\cite{4-1,4-2,4-3}, cold atom~\cite{5-1,5-2,5-3,5-4,5-5,5-6}, solid states~\cite{6-1,6-2}, etc. More recently, topological crystalline insulators (TCIs) hosting higher-order topological (HOT) states have drawn extensive attention~\cite{7-1,7-2,7-3,7-4}. Directly from the physical phenomena, for example, a two-dimensional (2D) second-order TCI can exhibit both one-dimensional (1D) gapped edge states and zero-dimensional (0D) corner states at mid-gap.
	
	However, hermiticity is only an ideal assumption of quantum mechanics. Recently, the study of topological states has gradually turned to open quantum system with non-Hermiticity~\cite{9-1}, which is closer to the real system. Many research have confirmed the topology of non-Hermitian systems, such as the photon and phono platform~\cite{10-1,10-2,10-3,10-4,10-5,10-6,10-7,10-8,10-9,10-10}, the cold atomic systems with gain and loss acting as non-Hermitian factor~\cite{11-1,11-2,11-3,11-4,11-5,11-6}, as well as the periodic driven Floquet non-Hermitian systems~\cite{12-1,12-2,12-3}.
	
	Generally, in non-Hermitian systems, the real eigenvalues are not obtained, and the eigen wavefunctions are no longer orthogonal. Instead, the left and right eigen wavefunctions with biorthogonality are needed, and some unique properties will appear, such as skin effect, exceptional point/ring, bulk Fermi arcs and so on~\cite{13-1,13-2,13-3}. In addition, for any total-loss non-Hermitian Hamiltonian, the gain-loss balanced part can be separated, whereby the PT symmetry of non-Hermitian total-loss system can be discussed.
	
	For the higher-order topology in non-Hermitian systems, it has been proved that the generalized bulk-corner correspondence can work for a wide class of non-Hermitian high-order topological systems with reflection or chiral symmetries. However, the existing non-Hermitian high-order topological insulators (NH-HOTIs) can not meet the diverse requirements of the real world, so it is of great significance to develop new NH-HOTIs. In addition, most of the existing studies consider the system with balanced gain and loss. But in reality, the laborious gain realization is especially unfavorable to the practical experimental research and application preparation. In this paper, we propose a NH-HOTI, the energy gap of which is opened by configuring the losses, and the edge and corner states in the energy gap are observed. Then we demonstrate the 'real-energy' edge states of pseudo-PT symmetry in total-loss non-Hermitian systems, and realize this pseudo-PT symmetry in NH-HOTIs domain wall system.

	

	\section{NH-HOTI}\label{2}
	In this section, we first briefly analyze the topological phase transition in two cases of loss configuration of 1D non-Hermitian SSH chain, then propose a NH-HOTI.

\begin{figure}
		\centering
		\includegraphics[width=8cm]{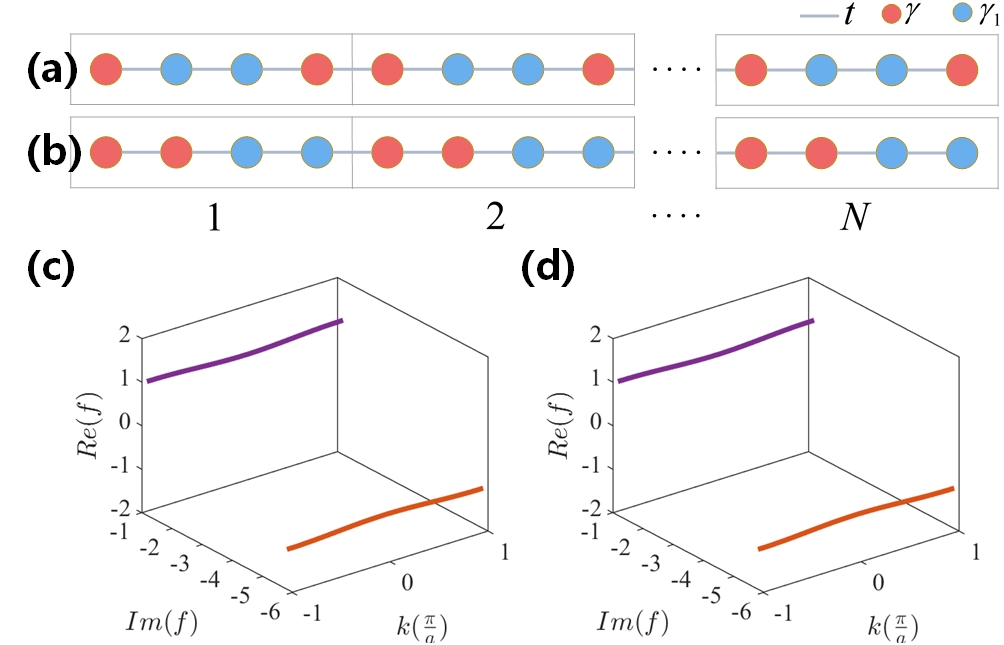}
		\caption{Schematic of 1D non-Hermitian SSH chain. (a) In case 1, the low-loss is on both sides, and the high-loss is added in the middle two cavities.(b) In case 2, there are two continuous background loss and additional loss in the unit cell. $N = 20$. The coupling strength between resonators is $t$. low-loss $\gamma$ and high-loss $\gamma_{1}$ are shown in red and blue, respectively. Energy spectrum of the model. The energy spectra of case 1 and case 2 are (c) and (d) respectively. The parameters of the both case: $t = 1$,$\gamma = 1$,$\gamma' = |\gamma''| = 5$.}
		\label{fig:one}
	\end{figure}

	We focus on non-Hermitian system of total-loss, and start with a regular 1D SSH chain consisting of N unit cells as depicted in Fig.~\ref{fig:one}. Here, this 1D SSH chain consist of four uniformly coupled cavities, with low-loss $\gamma$  (only background loss) introduced into two cavities and high-loss  $\gamma_{1}$ (background loss  $\gamma$ + additional loss $\gamma'$) introduced into the other two. The Hamiltonian of this chain based on the tight-binding (TB) model can be written as
	\begin{equation}
		\left(\begin{array}{cccc}
			-i\gamma&t&0&t\cdot e^{-iak} \cr
			t&-i(\gamma+\frac{\gamma'-\gamma''}{2})&t&0 \cr
			0&t&-i(\gamma+\gamma')&t \cr
			t\cdot e^{iak}&0&t&-i(\gamma+\frac{\gamma^{\prime}+\gamma^{\prime\prime}}{2})
			\label{eq:one}
		\end{array}	\right)
	\end{equation}
	
	where $a$ is the lattice constant and $k$ is the Bloch wave number. $t$ is the coupling strength between cavities. $\gamma'$and$\gamma''$ mark the loss and satisfy: $\gamma''=\pm\gamma'$. The imaginary onsite potential becomes $[-i\gamma,-i(\gamma+\gamma'),-i(\gamma+\gamma'),-i\gamma]$ where $\gamma' = -\gamma''$, which is called case 1, corresponding to Fig.~\ref{fig:one}(a). The imaginary onsite potential becomes $[-i\gamma,-i\gamma,-i(\gamma+\gamma'),-i(\gamma+\gamma')]$ where $\gamma' = \gamma''$, which is called case 2, corresponding to Fig.~\ref{fig:one}(b).The nontrivial topology of this system can be captured by a biorthogonal polarization~\cite{10-10,14-2,14-3}
	\begin{equation}
		P=\frac{i}{2\pi}\int_{0}^{2\pi}Tr[A(k)]\,dk
		\label{eq:two}
	\end{equation}
	where  $A(k)=\left\langle\phi_{\omega}(k)|\partial_{k}|\varphi_{\upsilon}(k')\right\rangle$ is the biorthogonal non-Ablien Berry connection. $|\phi_{\omega}(k)\rangle$,$|\varphi_{\upsilon}(k')\rangle $ represent left and right eigenvectors, and $\omega$ and $\upsilon$ denote the lower two occupied bands. The normalization condition $\left\langle\phi_{\omega}(k)|\varphi_{\upsilon}(k')\right\rangle=\delta_{\omega\upsilon}\delta_{kk'}$ is guaranteed. In general, the biorthogonal polarization is complex. Thus, the real-valued polarization can be defined as $P=P^{LR}+P^{RL}$. The biorthogonal polarization  can be calculated:$P=0.5$ in case 1 and $P=0$ in case 2, represent trivial topology and nontrivial topology, respectively.
	
\begin{figure}
		\centering
		\includegraphics[width=8cm]{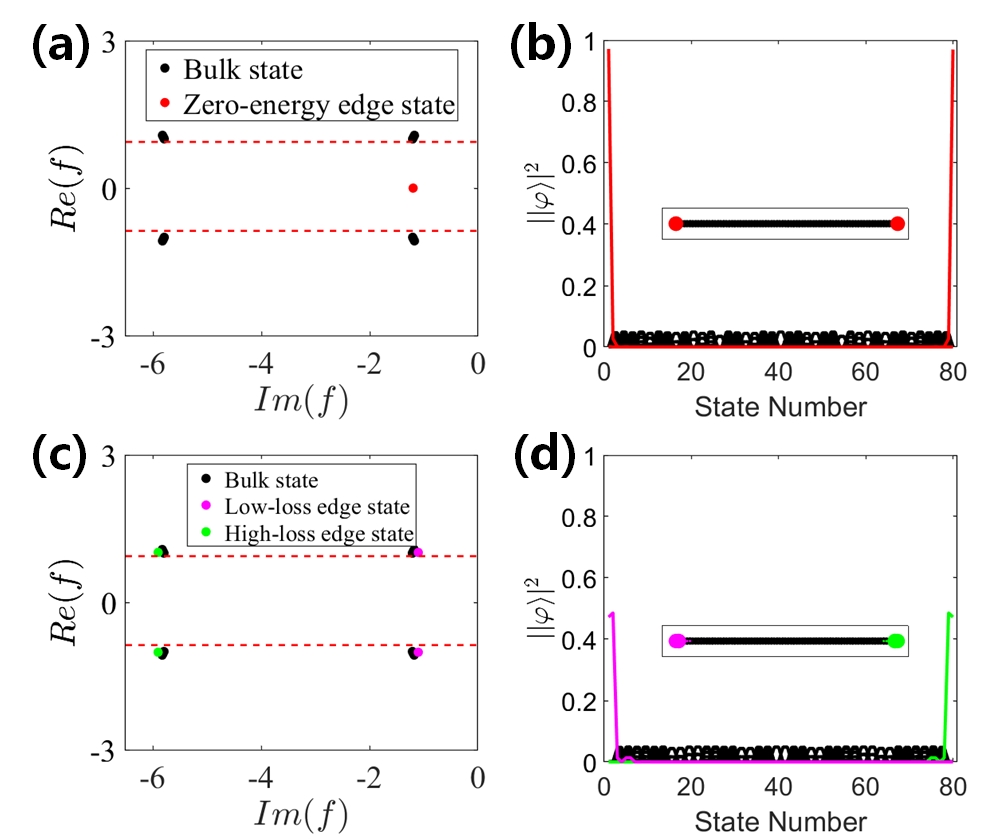}
		\caption{The bulk states (black), zero-energy edge states (red), low-loss edge states (magenta)  and high-loss edge states (green) are shown in (a)(b) (case 1) and (c)(d) (case 2), which is the eigenfrequency diagram of finite model with $N = 20$, and the distribution of eigenmode is shown in the illustration. All parameters are consistent with Fig.~\ref{fig:one}.}
		\label{fig:two}
	\end{figure}

	The topological property of this model can also be reflected in the energy band which opens a gap by the loss $\gamma'$ and $\gamma''$, and the energy spectrum are shown in Fig.~\ref{fig:one}(c)(d). We numerically calculate the eigenfrequency of the finite chain with open boundary on both sides, and get the result in Fig.~\ref{fig:two}. In case 1, we obtain the inner-gap edge states with zero-energy. The states here are very well localized on both sides.  In case 2, there is no inner-gap states, but two kinds of inner-band edge states appear, corresponding to low loss and high loss respectively. This shows that the two different loss configurations are trivial and nontrivial topology.

	Next, we will demonstrate the edge and corner states in a total-loss NH-HOTI. Here, we only consider the nearest-neighbor hopping, and neglect the real part of the onsite energy. The Hamiltonian of the corresponding TB model then can be given as:

	\begin{equation}
		H=
		\left(\begin{array}{cc}
			H_{1}&H_{2}\cr H_{3}&H_{4}
		\end{array}	\right)
	\end{equation}
	where
	
	\begin{widetext}
		\centering
		\begin{equation}
			H_{1}=
			\left(\begin{array}{cccccc}
				0&0&p_{11}&0&p_{21}&0  \cr  0&0&0&p_{21}&0&p_{31} \cr p_{12}&0&0&0&p_{31}&0 \cr 0&p_{22}&0&0&0&p_{21} \cr p_{22}&0&p_{32}&0&0&0 \cr 0&p_{32}&0&p_{22}&0&0
			\end{array}	\right)
			-\gamma\cdot
			\left(\begin{array}{cccccc}
				1&0&0&0&0&0 \cr 0&1&0&0&0&0 \cr 0&0&1&0&0&0 \cr 0&0&0&1&0&0 \cr 0&0&0&0&1&0 \cr 0&0&0&0&0&1
			\end{array}	\right); \notag
			H_{2}=
			t\cdot
			\left(\begin{array}{cccccc}
				1&0&0&0&0&1 \cr 1&1&0&0&0&0 \cr 0&1&1&0&0&0 \cr 0&0&1&1&0&0 \cr 0&0&0&1&1&0 \cr 0&0&0&0&1&1
			\end{array}	\right) \notag
			\label{eq:five}
		\end{equation}
		\centering
		\begin{equation}
			H_{3}=
			t\cdot
			\left(\begin{array}{cccccc}
				1&1&0&0&0&0 \cr 0&1&1&0&0&0 \cr 0&0&1&1&0&0 \cr 0&0&0&1&1&0 \cr 0&0&0&0&1&1 \cr 1&0&0&0&0&1
			\end{array}	\right); \notag
			H_{4}=t\cdot
			\left(\begin{array}{cccccc}
				0&1&0&0&0&1 \cr 0&0&1&0&0&0 \cr 0&0&0&1&0&0 \cr 0&0&0&0&1&0 \cr 0&0&0&0&0&1 \cr 0&0&0&0&0&0
			\end{array}	\right)
			-(\gamma+\gamma')\cdot
			\left(\begin{array}{cccccc}
				1&0&0&0&0&0 \cr 0&1&0&0&0&0 \cr 0&0&1&0&0&0 \cr 0&0&0&1&0&0 \cr 0&0&0&0&1&0 \cr 0&0&0&0&0&1
			\end{array}	\right)\notag
			\label{eq:four}
		\end{equation}
	\end{widetext}

\begin{figure}
		\centering
		\includegraphics[width=8cm]{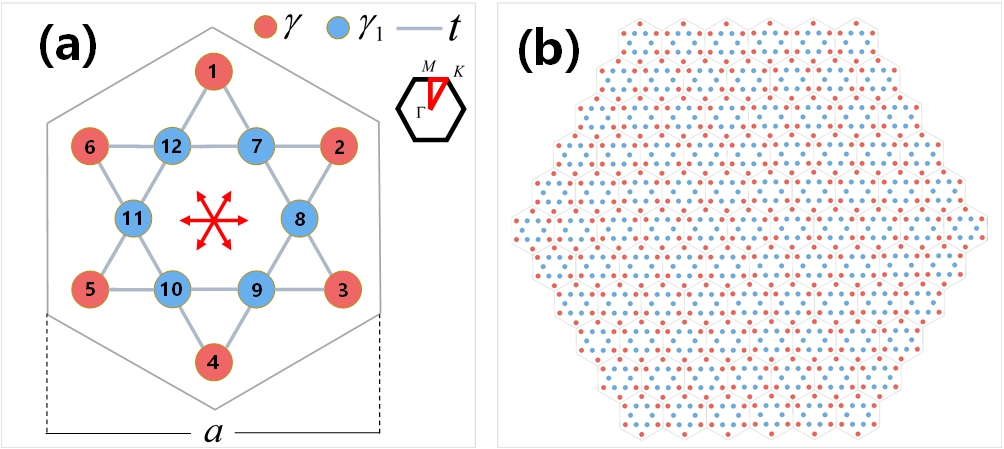}
		\caption{(a) 12 resonant cavities and the loss configuration(case 3). Red and blue represent two different lattice points of low-loss and high-loss respectively, $a$ is the lattice constant of the unit cell. Red arrows mark nontrivial topology in six directions. The illustration is the first Brillouin zone. As the model satisfies the C6 symmetry, the simplified Brillouin zone is described by the red triangle. (b) is a finite model based on (a).}
		\label{fig:six}
	\end{figure}

	where $p_{11} = t\cdot exp(i\cdot (-\frac {1}{2}k_{x}+\frac {\sqrt{3}}{2}k_{y}))$;$p_{21} = t\cdot exp(i\cdot (\frac {1}{2}k_{x}+\frac {\sqrt{3}}{2}k_{y}))$;$p_{31} = t\cdot exp(i\cdot k_{x})$;$p_{12} = p^{\ast}_{11}$;$p_{22} = p^{\ast}_{21}$;$p_{32} = p^{\ast}_{31}$ are the hopping unit vectors. $\ast$ denotes complex conjugate. The value of $t$, $\gamma$ and $\gamma^{\prime}$ is consistent with the previous. As is shown in Fig.~\ref{fig:six}(a), the unit cell is composed of six internal (Armchair) and external (Zigzag) resonant cavities. The six external cavities only have low-loss $\gamma$, and the six internal cavities have high-loss(case 3). This case can be regarded as case 1 of 1D SSH chain in the first section in six directions and demonstrated in Sec III.

\begin{figure*}
		\centering
		\includegraphics[width=16.5cm]{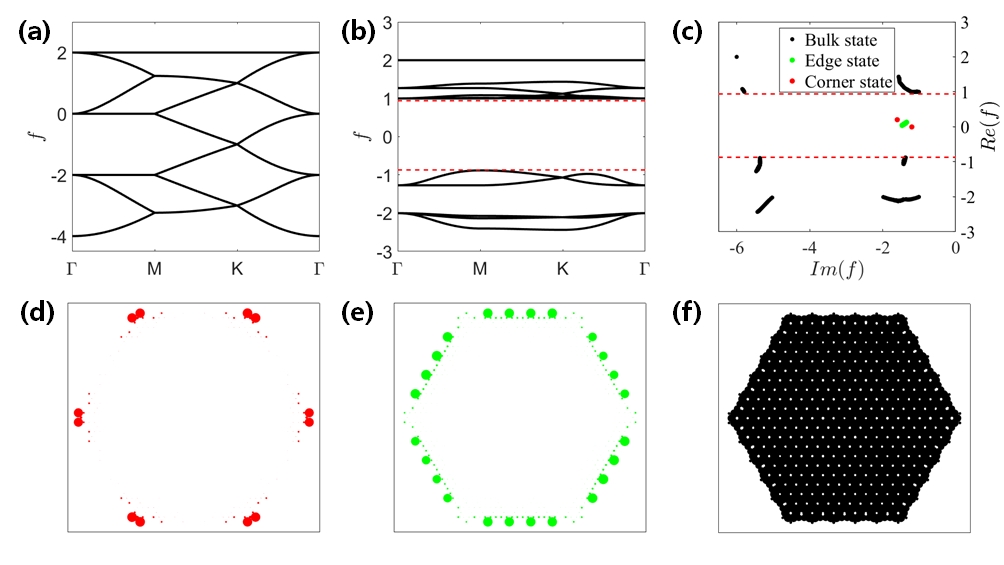}
		\caption{(a) Energy band of lossless system($\gamma = 1,\gamma_{1} = 6$). (b) Energy band with two kinds of loss. The red dotted line depicts the topological band gap. (c) Eigenfrequency spectrum of finite lattice model in Fig.~\ref{fig:six}(b). Abscissa and ordinate represent the real and imaginary part respectively. The black dot, green dot and red dot represent the bulk states, edge states and corner states, respectively. Eigenmodes (right eigenvector) of finite lattice: (d) corner states (e) edge states (f) bulk states.}
		\label{fig:seven}
	\end{figure*}

	Then we calculate the band structure along the $\Gamma-M-K-\Gamma$ path. The cases of lossless and loss-considering are shown in Fig.~\ref{fig:seven}(a) and (b), respectively. Three bandgaps emerge after two different losses are introduced. In addition, we numerically calculate the eigenfrequency of the finite 2D modelFig.~\ref{fig:seven}(c). In addition to the bulk states, we also find the inner-gap edge states and corner states which are the only two discussed states here. We plot the sum of probabilities (using right eigenvectors) of the bulk, edge, and corner states in Fig.~\ref{fig:seven}(d–f), respectively, further identifying their existence. We note, as can be seen from Fig.~\ref{fig:seven}(b–d), the states distribute all over the bulk, edges, and corners, so this system does not feature non-Hermitian skin effect, but belongs to higher-order topological system.  
	
	Here, we calculate the wannier band in case 3, to characterize its topology.
	First, we use $H|\varphi\rangle=E_{1}|\varphi\rangle$, $H^{\dagger}|\phi\rangle=E_{2}|\phi\rangle$ to define the right $|\varphi\rangle$ and left $|\phi\rangle$ eigenstate. The Brillouin zone is discretized into $M \times N$ k-points along $b_{1}$ and $b_{2}$ direction respectively~\cite{SM}(we use $M = N = 100$ in all calculations). Then we define Wilson loop element: $\Re_{\omega,\upsilon}(k) = \left\langle\phi_{\omega}(k)|\partial_{k}|\varphi_{\upsilon}(k+\triangle k)\right\rangle$ in which $\triangle k=\frac{2\pi}{N}$. Now, the wilson loop can be defined:
	\begin{equation}
		W_{\omega,\upsilon}=F_{\omega,\upsilon}(k) \cdot F_{\omega,\upsilon}(k+\triangle k) \cdot \cdot \cdot F_{\omega,\upsilon}(k+(N-1)\triangle k)
		\label{eq:seven}
	\end{equation}
	where 
	\begin{equation}
		F_{\omega,\upsilon}(k)=
		\left(\begin{array}{cccc}
			\Re_{1,1}(k)&\Re_{1,2}(k)&\cdots&\Re_{1,\upsilon}(k)  \cr 
			\Re_{2,1}(k)&\Re_{2,2}(k)&\cdots&\Re_{2,\upsilon}(k)  \cr 
			\vdots&\vdots&\ddots&\vdots  \cr 
			\Re_{\omega,1}(k)&\Re_{\omega,2}(k)&\cdots&\Re_{\omega,\upsilon}(k))
			\label{eq:eight}
		\end{array}	\right)
	\end{equation}

\begin{figure}
		\centering
		\includegraphics[width=8cm]{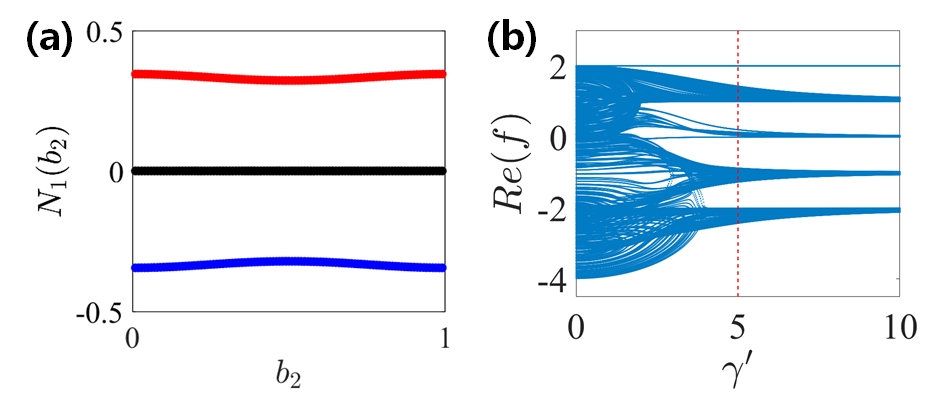}
		\caption{(a) Wannier band in case 3(red, blue and black are used to mark wannier center $>0$, $<0$ and $=0$, respectively). (b) Calculated eigenfrequencies for Fig.~\ref{fig:six}(b) as a function of the $\gamma'$(case 3). The red dashed line indicates the parameter value used in Fig.~\ref{fig:seven}.}
		\label{fig:twelve}
	\end{figure}

	We then diagonalize the Wilson loop operator $W_{\omega,\upsilon}|\chi\rangle = exp(i\cdot 2\pi N_{1}(b_{2}))|\chi\rangle$, where $|\chi\rangle$ is the eigenvector which depends on the wilson loop, and the phase $N_{1}(b_{2})$ is the wannier center. The wannier band is calculated(Fig.~\ref{fig:twelve}(a)) by using five bands under the second band gap, that is, $\omega$ and $\upsilon$ are equal to 5. It can be seen that the calculated Wannier band is gapped, which is the reason why edge states and corner states appear in the second band gap~\cite{14-2}.

\begin{figure}
		\centering
		\includegraphics[width=8cm]{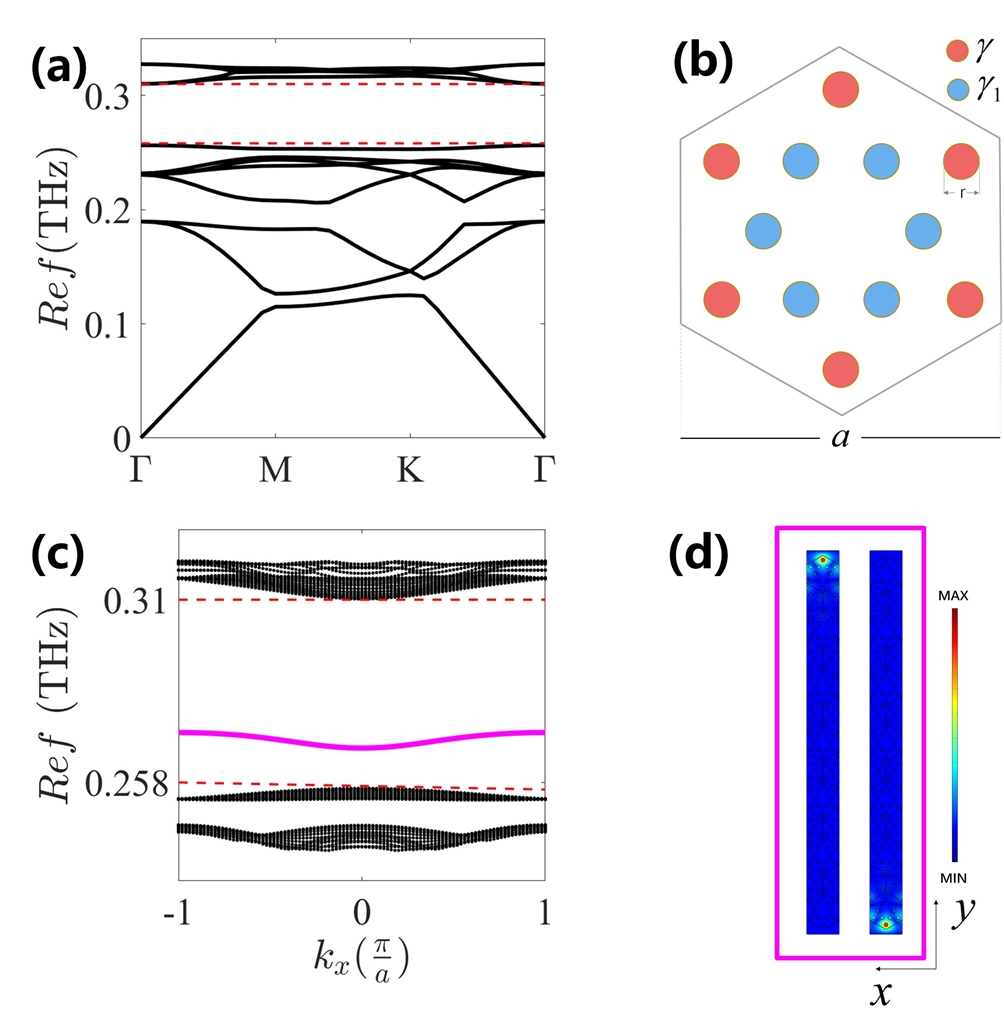}
		\caption{Band structure and schematics of non-Hermitian optical photonic structure. (a) and (c) are band structures of unit cell and supercell, respectively. The red dotted line marks the topological energy gap, and the magenta solid line is the edge states. (b) The lattice constant $a = 800 um$, the lattice radius $r = 0.12 a$, the real part of the relative permittivity of the material $\epsilon_{Re} = 11.7$, the imaginary part of the background loss $\epsilon_{Im} = 1$, and after adding loss $\epsilon_{Im} = 15$, and (d) is the eigenmodes of the simulated supercell edge states.}
		\label{fig:nine}
	\end{figure}

	Next, we realize the NH-HOTI through the photonic crystal platform. The energy band structure and the unit cell are shown in Fig.~\ref{fig:nine}(a)(b). Here, the loss is introduced through the imaginary part of the non-zero imaginary part of relative permittivity. We simulate the supercell with finite $y$ direction and obtain the energy band structure of the supercell with edge states (as shown in Fig.~\ref{fig:nine}(c)). Without the domain wall, supercell has edge states as shown in Fig.~\ref{fig:nine}(d), with two edge states on two boundaries.

\begin{figure}
		\centering
		\includegraphics[width=8.5cm]{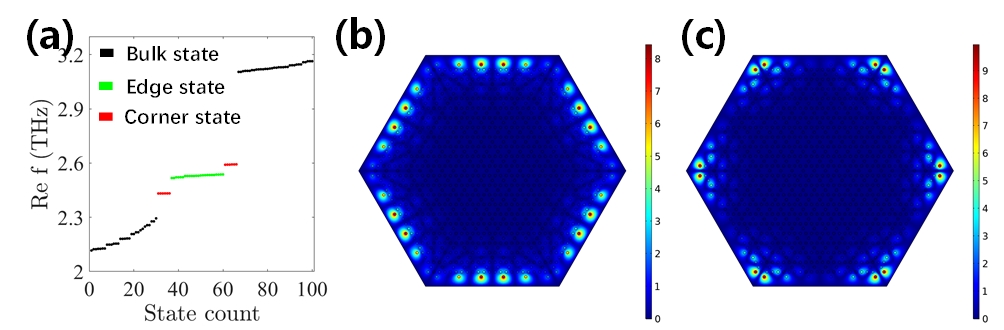}
		\caption{Simulation results of finite lattice model. (a) is the eigenfrequency of hexagonal finite lattice model. (b) (c) are the edge and corner mode field distributions of the green and red dots in (a).}
		\label{fig:ten}
	\end{figure}

	Then we simulate the eigenfrequency of the finite hexagonal lattice model, as is shown in Fig.~\ref{fig:ten}(a). The corresponding edge states appear at the frequency corresponding to the supercell band structure (shown by the green and red dots in the figure).

	\section{Pseudo-PT symmetric in total-loss non-Hermitian system}\label{3}
	
	In this section, we will demonstrate that there exist 'real-energy' edge states in total-loss non-Hermitian pseudo-PT symmetric systems, which are implemented in 1D SSH chain and NH-HOTIS syetems. Then, the flexible correspondence between 1D non-Hermitian SSH chain and NH-HOTI is discussed.
	
	Generally, in a non-Hermitian system with gain and loss, a domain wall system satisfying PT symmetry can be constructed by configuring the size of loss and gain~\cite{14-1}. This system grows 'real-energy' edge states embedded in the bulk states. Here we find that there also exist 'real-energy' edge states embedded in the bulk states in the total-loss non-Hermitian system.
	
	For the total-loss non-Hermitian pseudo-PT symmetric domain wall system, we define: the whole domain wall system can balance the gain and loss around a 'real-energy'. In addition, the generalized PT part can be separated from the pseudo-PT symmetric system based on the 'real-energy'. Also satisfy $PTH_{PT}^{L} = H_{PT}^{R}$, where $H_{PT}^{L}, H_{PT}^{R}$ are the Hamiltonian of left and right sides of the domain wall which separated from the pseudo-PT symmetric system. Here, we use case 1 in Section II to construct a pseudo-PT domain wall system for example.
	
	For any total-loss non-Hermitian system, the PT symmetric part can be separated. In case 1, the Hamiltonian can be written as $H = H_{PT}+H'=$
	
	\begin{equation}
		\left(\begin{array}{cccc}
			
			i\frac{\gamma'}{2}&t&0& e^{-iak} \cr 1&-i\frac{\gamma'}{2}&1&0 \cr 0&1&-i\frac{\gamma'}{2}&1 \cr e^{iak}&0&1&i\frac{\gamma'}{2}
		\end{array}	\right)
		-
		i\gamma_{Re}\cdot
		\left(\begin{array}{cccc}
			1&0&0&0 \cr  0&1&0&0 \cr 0&0&1&0 \cr 0&0&0&1
		\end{array}	\right)
		\label{eq:three}
	\end{equation}
	
	where $-i\gamma_{Re} = -i(\gamma+\frac{\gamma'}{2})$ is the 'real-energy'.

\begin{figure}
		\centering
		\includegraphics[width=8cm]{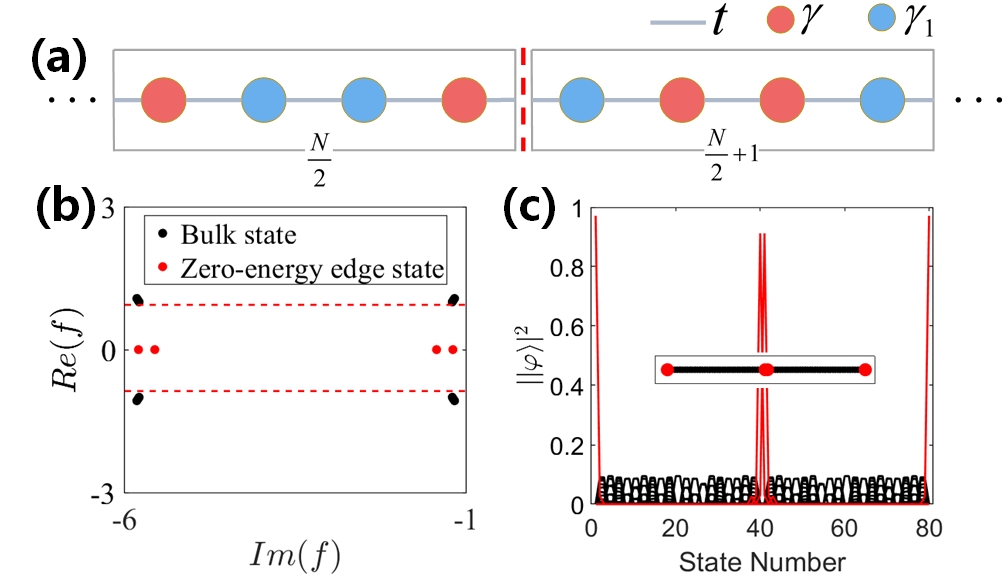}
		\caption{The Schematic, eigenfrequencies and eigenmodes of pseudo-PT symmetric domain wall system. (a) is the finite model of 1D SSH chain in case 1 with $N = 20$. At both sides of the domain wall, the positions of the imaginary parts $\gamma$ and $\gamma_{1}$ of the potential energy of the cavity are reversed. (b) is the eigenfrequencies distribution. (c) is the eigenmodes distribution. It can be seen that the edge states appear at both the domain wall and the side in the same way as the case without domain wall, that is, only corresponding to the loss. The meaning representative of colors and the selection of parameters are consistent with the Fig.~\ref{fig:two}.}
		\label{fig:three}
	\end{figure}

	According to this concept, we can construct pseudo-PT symmetric domain wall system in 1D SSH chain. First, we show a pseudo-PT symmetric system without a 'real-energy' edge states. Based on the non-Hermitian model of case 1 and case 2, we construct it by creating domain wall in the middle of the whole model, as shown in Fig.~\ref{fig:three}(a). The numerical results (Fig.~\ref{fig:three}(b)(c)) show that there are zero-energy edge states at the domain wall. The appearance of edge states is a simple combination of two loss configurations, means that the imaginary part of the frequency of edge states at the domain wall corresponds to the size of loss at the domain wall.

\begin{figure}
		\centering
		\includegraphics[width=8cm]{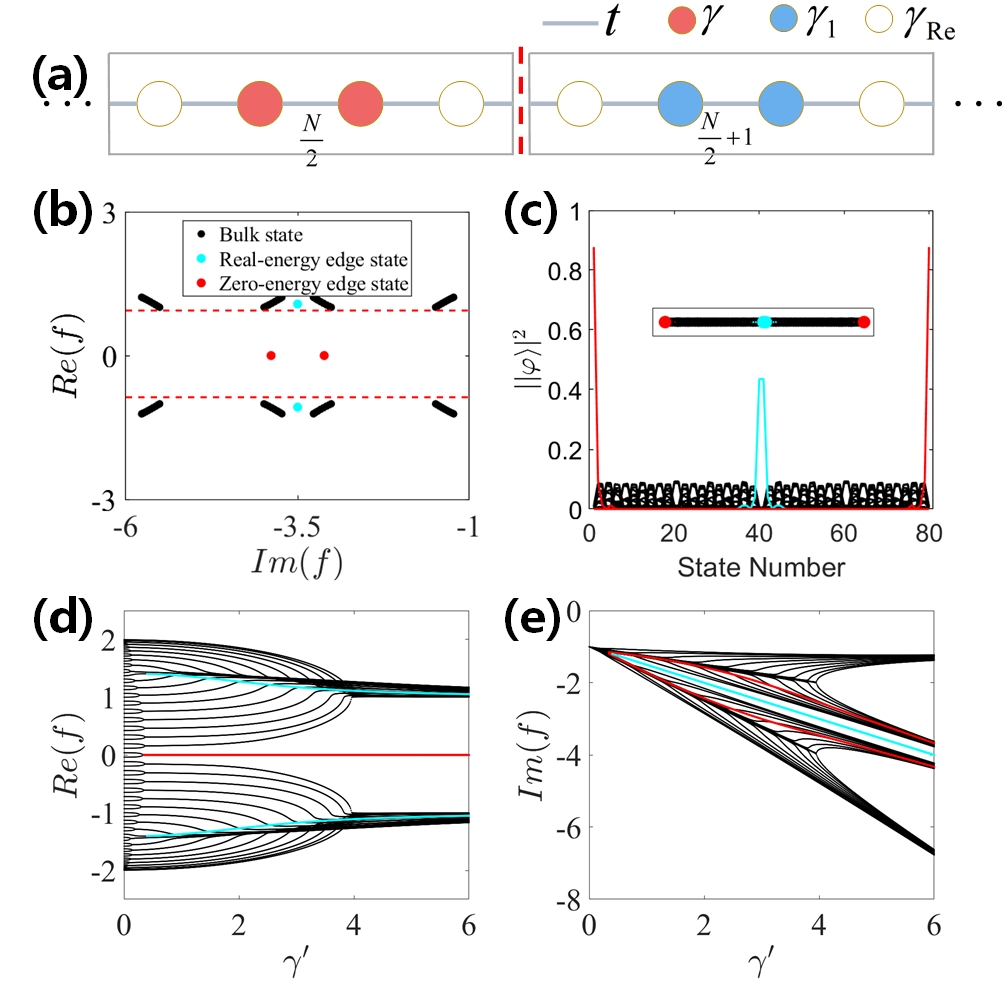}
		\caption{The schematic, eigenfrequencies and eigenmodes of pseudo-PT symmetric domain wall system. (a) is the finite model of SSH chain with $N = 20$. Representative of low-loss and high-loss is same as before, and white dots indicate loss of $\gamma_{Re}$. (b) and (c) are the eigenfrequencies and eigenmodes of this pseudo-PT symmetric domain wall system, respectively. (d) and (e) are the real part and imaginary part of eigenfrequencies as a function of the $\gamma'$. The meaning representative of colors and the selection of parameters are consistent with the Fig.~\ref{fig:two}.}
		\label{fig:four}
	\end{figure}

	According to Eq.~(\ref{eq:three}), in the pseudo-PT symmetric domain wall system with 'real-energy' edge states, the Hamiltonian can have a 'real-energy' part in addition to the PT symmetric part. We construct the domain wall according to this 'real-energy', and the imaginary parts of the potential energy on both sides of the domain wall are $[-i\gamma_{Re},-i\gamma,-i\gamma,-i\gamma_{Re}]$ and $[-i\gamma_{Re},-i\gamma_{1},-i\gamma_{1},-i\gamma_{Re}]$. In the numerical calculation of the domain wall system with $\gamma_{Re} = 3.5$(for $\gamma=1$ and  $\gamma'=5$), it is obtained that the 'real-energy' edge states in the bulk states appears at the imaginary part of the eigenfrequency of -3.5 (Fig.~\ref{fig:four}(d)). In addition, the domain wall system also has the edge states corresponding to the size of loss at both sides. In Fig.~\ref{fig:four}(e), it can be seen that the imaginary part of the characteristic frequency is linear with $\gamma'$.

	Then, we use this NH-HOTI which can be regarded as 1D non-Hermitian SSH chain in six directions to fabricate a 2D pseudo-PT symmetric domain wall system, and obtain the 'real-energy' edge states at the domain wall.
	
\begin{figure}
		\centering
		\includegraphics[width=8cm]{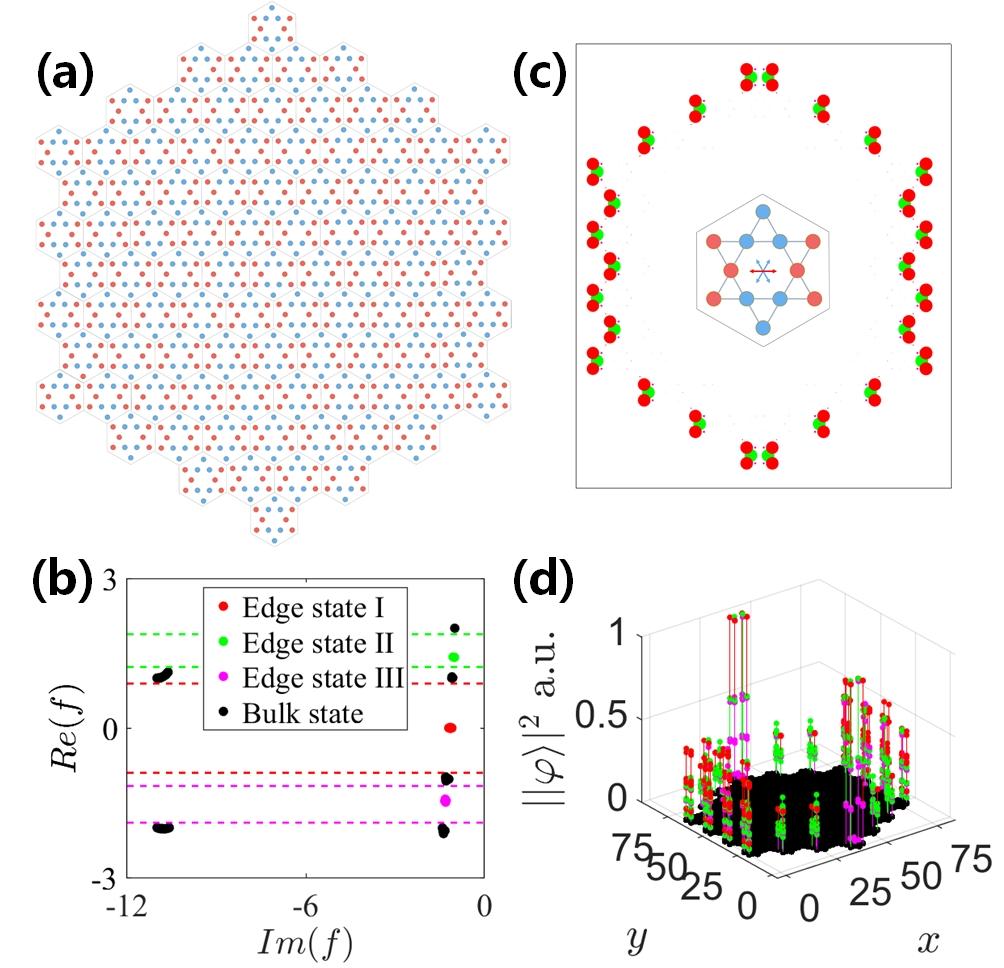}
		\caption{(a) Finite model based on the case: add low-loss at '2 3 5 6 8 11' and add high-loss at other positions($\gamma = 1$, $\gamma_{1} = 11$). Its eigenfrequency spectrum is (b) in which edge state I(red dots), edge state II(green dots), edge state III(magenta dots) exist in three band gaps(marked with dashed lines of three corresponding colors) respectively, and the loss configuration is called case 4. (c) is the eigenmodes of (a), and the unit cell is shown in the illustration in which red and blue arrows mark nontrivial and trivial topology in different directions. (d) is the sum of the square of right eigenvertors module corresponding to edge states I, II and III in (b).}
		\label{fig:eleven}
	\end{figure}

	In order to verify that NH-HOTI can be regarded as a simple superposition of 1D SSH chain in Sec II in six different directions, we use a loss configuration different from Sec II: add low-loss at '2 3 5 6 8 11' and add high-loss at other positions(case 4). In this instance, it is a nontrivial topology only in the positive and negative directions of x, and we construct a finite model(Fig.~\ref{fig:eleven}(a)) based on this loss configuration. The eigenmodes of three edge states show that the required states only appears in the positive and negative directions of x, which is exactly what we want and need. Wannier bands of three bands with edge states are shown in the appendix~\cite{SM}.

\begin{figure}
		\centering
		\includegraphics[width=8cm]{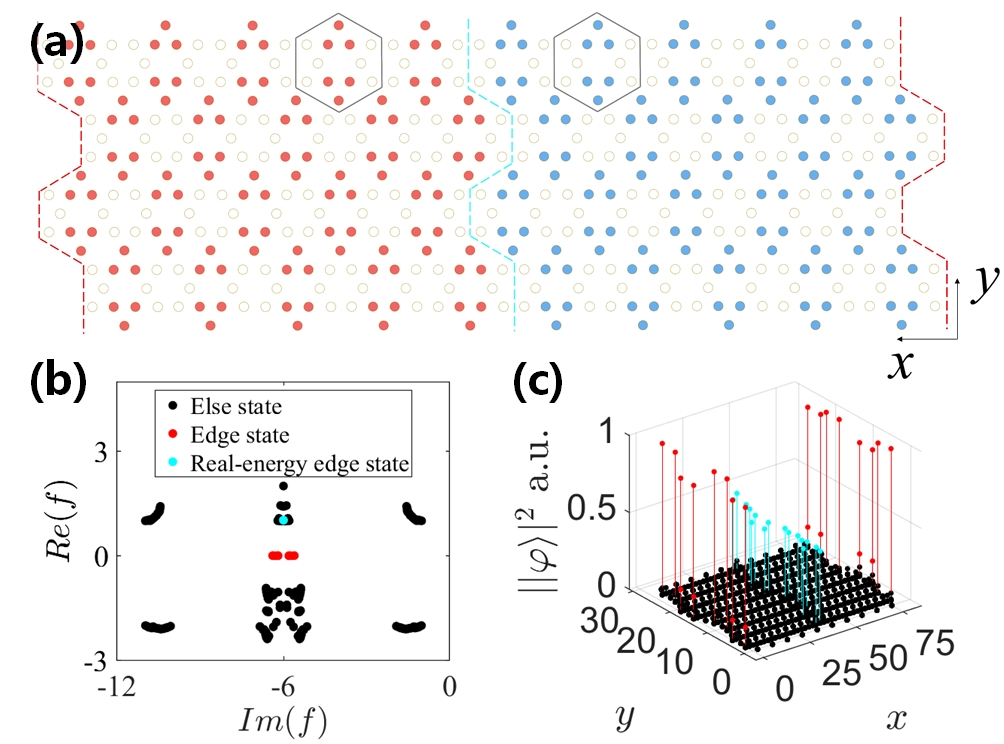}
		\caption{The schematic, eigenfrequencies and eigenmodes of pseudo-PT symmetric domain wall system. (a) is the finite model constructed of 4*10 unit cells. The imaginary onsite potential of red, white and blue: $\gamma = 1$, $\gamma_{Re} = 6$, $\gamma_{1} = 11$, respectively. The black solid line marks the unit cell, and the cyan and red dotted lines mark the domain wall and both end walls. (b) is the eigenfrequencies of this system, in which, red and cyan dots are the edge states at the end wall and the 'real-energy' edge states at the domain wall, and black dots are states other than the edge states marked in cyan and red. (c) is the sum of the square of right eigenvertors module corresponding to edge states and 'real-energy' edge states in (b).}
		\label{fig:eight}
	\end{figure}

	Next, wo add the loss as $i\gamma_{Re}$ at '2 3 5 6 8 11' and add low-loss and high-loss on two sides at other positions respectively. In this way, we obtain the loss configuration shown in Fig.~\ref{fig:eight}(a), which is the case as Fig.~\ref{fig:four}(a) in $x$ direction. Then we construct a finite domain wall system of this model, which is composed of 4 * 10 unit cells. This domain wall system only has edge states at both ends of the $x$ direction of the whole system, except for forming 'real-energy' boundary states at the domain wall, as is shown in Fig.~\ref{fig:eight}(b).
	
	So far, we have demonstrated that 1D SSH chain with non-trivial or trivial topology can be constructed in different directions, and various high-order topology models that meet different needs can be obtained flexibly.
	
	\section{CONCLUSION}\label{4}
	
	In summary, we propose a total-loss NH-HOTI, which realizes the existence of edge states and corner states in the energy band gap by adjusting the loss configuration, and can obtain edge states in different directions by adjusting the loss configuration flexibly. The topology of this NH-HOTI can be characterized by the gapped wannier band. In addition, we discuss the existence of 'real-energy' edge states in pseudo-PT symmetric domain wall system, and demonstrate it theoretically in the domain wall system of 1D SSH chain and NH-HOTI. Finally, we demonstrate the flexible correspondence between 1D non-Hermitian SSH chain and NH-HOTI. These results are expected to be useful for exploring high-order non-Hermitian models, and provide a flexible and useful scheme for finding topological states of real systems considering loss only.

	\begin{acknowledgments}
		Acknowledgments: This work was supported by the National Natural Science Foundation of China (Grant numbers 61865009, 61927813).
	\end{acknowledgments}




\end{document}